\documentclass[11pt]{revtex4}
\usepackage{amsfonts}
\usepackage{amssymb,epsf}
\usepackage{latexsym}                                  

\begin{document}
\title{Scalar Field Reconstruction of  Power-Law Entropy-Corrected HDE}

\author{Esmaeil Ebrahimi$^{1,2}$ \footnote{eebrahimi@uk.ac.ir} and Ahmad  Sheykhi$^{2,3}$ \footnote{
sheykhi@uk.ac.ir} }
\address{$^1$Department of Physics, Shahid Bahonar University, PO Box 76175, Kerman, Iran\\
          $^2$ Research Institute for Astronomy and Astrophysics of Maragha (RIAAM), Maragha, Iran\\
          $^3$ Physics Department and Biruni Observatory, Shiraz University, Shiraz 71454,
          Iran}

\begin{abstract}
\vspace*{1.5cm} \centerline{\bf Abstract} \vspace*{1cm} A so
called ``power-law entropy-corrected holographic dark energy"
(PLECHDE) was recently proposed to explain the dark energy
dominated universe. This model is based on the power-law
corrections to black hole entropy which appear in dealing with the
entanglement of quantum fields between inside and outside of the
horizon. In this paper, we suggest a correspondence between
interacting PLECHDE and tachyon, quintessence, K-essence and
dilaton scalar field models of dark energy in a non-flat FRW
universe. Then, we reconstruct the potential terms accordingly,
and present the dynamical equations which describe the evolution
of the scalar field dark energy models.
\end{abstract}

 \maketitle

\newpage
\section{Introduction}
One of the most dramatic field of research in theoretical physics,
these days, is the investigation on the cause of an unpredicted
phase of accelerated expansion in our universe. According to
different cosmological observations, our universe is undergoing a
phase of accelerated expansion which the cause of it has not been
known yet \cite{1,2,3,4,5}. A component which is responsible for
this acceleration usually called dark energy (DE). Most of the
observations confirm that the DE consists more than $70\%$ of the
energy content of our universe and the nature of such a component is
still unknown \cite{padmanabhan}. The simplest candidate for DE is
the cosmological constant which leads to $w=-1$ \cite{sahni}.
Though, it suffers the so-called {fine-tuning} and {cosmic
coincidence} problems \cite{2}. However, observations detect a
deviation from the models with constant equation of state (EoS)
parameter and show an evolution in the EoS parameter. Due to this
fact another category of models has been proposed as possible
candidates for DE. In these models the DE candidate has a dynamical
behavior and lead to a variable EoS parameter. The simplest
candidate in the dynamic approach is the scalar field $\phi$. This
approach for probing the nature of DE has been extensively studied
in the literature. Some famous examples of these models are
quintessence, tachyon, K-essence, dilaton field and so on (see
\cite{wettrich,chiba,armenda1,armenda2} and references therein). For
a recent review on DE models see \cite{Pad}.

Among various attempts toward understanding the DE puzzle, the
holographic dark energy (HDE) model has got a lot of enthusiasm
recently. This model, which has been widely studied in the
literature \cite{Coh,Li,Huang,Hsu,pav1,Ricci,GO,Sad,wang1,HDE,wang2,Sheykhi1}, is
motivated from the holographic hypothesis \cite{Suss1} and has
been tested and constrained by various astronomical observations
\cite{Xin,Feng}. It is important to note that in the derivation of
HDE density the black hole entropy $S$ plays a crucial role.
Indeed, the definition and derivation of holographic energy
density ($\rho_{D}=3c^2M_p^2/L^2$) depends on the
entropy-area relation $S\propto A \propto L^2$ of black holes,
where $A$ represents the area of the horizon \cite{Coh}. However,
quantum corrections to the area law have been introduced in recent
years, namely, logarithmic and power-law corrections. Logarithmic
corrections, arises from loop quantum gravity due to thermal
equilibrium fluctuations and quantum fluctuations \cite{Meis},
\begin{equation}\label{celog}
    S=\frac{A}{4G}+\gamma \ln{\frac{A}{4G}}+\delta,
\end{equation}
where $\gamma$ and $\delta$ are dimensionless constants of order
unity. The exact values of these constants are not yet determined
and still an open issue in quantum gravity.  This logarithmic term
also appears in a model of entropic cosmology which unifies the
inflation and late time acceleration \cite{cai}. Another form of the
correction to the area law, namely the power-law correction, appears
in dealing with the entanglement of quantum fields in and out the
horizon. The entanglement entropy of the ground state obeys the
Hawking area law. Only the excited state contributes to the
correction, and more excitations produce more deviation from the
area law \cite{sau1} (also see \cite{sau2} for a review on the
origin of black hole entropy through entanglement). The power-law
corrected entropy is written as \cite{Sau,pavon1}
\begin{equation}\label{plec}
S=\frac{A}{4G}\left[1-K_{\alpha}A^{1-\alpha/2}\right],
\end{equation}
where $\alpha$ is a dimensionless constant whose value is
currently under debate, and
\begin{equation}\label{kalpha}
K_{\alpha}=\frac{\alpha(4\pi)^{\alpha/2-1}}{(4-\alpha)r_c^{2-\alpha}}.
\end{equation}
Here $r_c$ is the crossover scale. The second term in Eq. (2) can be
regarded as a power-law correction to the area law, resulting from
entanglement, when the wave-function of the field is chosen to be a
superposition of ground state and exited state \cite{Sau}.

The HDE models with logarithmic correction have been explored in
ample details \cite{ECHDE}. Very recently, by taking into account
the power-law correction to entropy, a so called ``PLECHDE" was
proposed by Sheykhi and Jamil \cite{Sheykhi2}. It was shown that
this model is capable to provide an accelerated expansion. Due to
the variety of models in the literature, to explain the DE problem,
it seems essential to develop a correspondence between different
approaches clarifying the theoretical status of the current models.
To this end, in this paper we would like to implement the
correspondence between scalar field enrgy density and PLECHDE model.
This connection allows us to reconstruct the potential as well as
the dynamics of the scalar fields which describe the acceleration of
the universe expansion.

This paper is outlined as follows. In the next section we
establish the correspondence between PLECHDE and the tachyon field
dark energy model. In section \ref{qerec}, we study the
quintessence dark energy model based on the interacting PLECHDE.
Section \ref{kessrec} consists the reconstruction procedure of the
K-essence PLECHDE. In section \ref{dilrec}, we consider the
reconstructed model of the dilatonic DE based on the PLECHDE. We
finish the paper with some concluding remarks in section
\ref{con}.

\section{Tachyon reconstruction of PLECHDE} \label{tachrec}
 The tachyon model of dark energy originates from string theory
 and seems to has interesting cosmological consequences.  One of the interesting properties of a rolling
 tachyon is its EoS parameter which changes between $-1$ to $0$ \cite{Gib1}.
 This EoS parameter convinces people to consider the tachyon scalar
 field as a candidate for DE. It was demonstrated that dark energy driven by
tachyon decays to cold dark matter in the late accelerated
universe and this phenomenon yields a solution to cosmic
coincidence problem \cite{Sri}. Choosing different
self-interacting potentials in the tachyon field model lead to
different consequences for the resulting DE model. Due to this
fact we would like to reconstruct the tachyon equivalent of the
PLECHDE to see which tachyon scalar field model can demonstrate
quantum gravity effects. The connection between tachyon field and
HDE \cite{Setare4}, agegraphic dark energy (ADE) \cite{ahmad},
ECHDE \cite{tachechde} and ECADE models \cite{ahmadechde} have
been already established. The effective lagrangian for the tachyon
field is described by
\begin{eqnarray}
 L=-V(\phi)\sqrt{1-g^{\mu\nu}\partial_\mu \phi \partial_\nu \phi},
 \end{eqnarray}
where $V(\phi)$ is the tachyon potential. The corresponding energy
momentum tensor for the tachyon field can be written in a perfect
fluid form
\begin{eqnarray}
 T_{\mu\nu}=(\rho_\phi+p_\phi)u_{\mu} u_\nu-p_\phi g_{\mu\nu},
 \end{eqnarray}
where $\rho_\phi$ and $p_\phi$ are, respectively, the energy
density and pressure of the tachyon, and the velocity $u_\mu$ is
\begin{eqnarray}
u_\mu=\frac{\partial_\mu \phi}{\sqrt{\partial_\nu \phi \partial^\nu
\phi}}.
 \end{eqnarray}
We assume the background Friedmann-Robertson-Walker (FRW) metric which
is described by the line element
\begin{equation}
ds^2=dt^2-a^2(t)\Big(\frac{dr^2}{1-kr^2}+r^2d\Omega^2\Big),
\end{equation}
where $a(t)$ is the scale factor,  $k$ is the curvature parameter
with $k = -1, 0, 1$ corresponding to open, flat, and closed
universes, respectively. The first Friedmann equation takes the form
\begin{eqnarray}\label{Fried}
H^2+\frac{k}{a^2}=\frac{1}{3M_p^2} \left( \rho_m+\rho_D \right),
\end{eqnarray}
where $\rho_m$ and $\rho_D$ are, respectively, the energy densities of pressureless matter and
dark energy. The dimensionless density parameters are defined as
usual
\begin{equation}
\Omega_m=\frac{\rho_m}{\rho_{cr}},\ \ \
\Omega_D=\frac{\rho_D}{\rho_{cr}},\ \ \Omega_k=\frac{k}{a^2H^2},
\end{equation}
where the critical energy density is $\rho_{cr}={3H^2 M_p^2}$.
Thus the first Friedmann equation can be rewritten as
\begin{equation}\label{fridomega}
1+\Omega_k=\Omega_m+\Omega_D.
\end{equation}
The energy density and pressure of tachyon field are given by
\begin{equation}
\rho_\phi=-T_0^0=\frac{V(\phi)}{\sqrt{1-\dot\phi^2}},\label{rhophi}
\end{equation}
\begin{equation}
p_\phi=T_i^i=-V(\phi)\sqrt{1-\dot\phi^2}.
\end{equation}
The equation of state parameter is
\begin{equation}
w_\phi=\frac{p_\phi}{\rho_\phi}=\dot\phi^2-1.
\end{equation}
Following \cite{Sheykhi2}, we assume the energy density of the
PLECHDE has the the following form
\begin{equation}\label{rhopl}
\rho_D=3c^2M_p^2L^{-2}-\beta M_p^2
L^{-\alpha}=\frac{3c^2M_p^2}{L^2}\gamma_c,
\end{equation}
where
\begin{equation}\label{gamma}
    \gamma_c=1-\frac{\beta}{3c^2L^{\alpha-2}},
\end{equation}
 and $L$ is a length scale which provides an IR cut-off
for the holographic model of dark energy. In the literature a
variety of IR cut-offs have been assumed. Li \cite{Li} discussed
three choices for the length scale $L$ which is supposed to
provide an IR cut-off. The first choice is the Hubble radius,
$L=H^{-1}$ \cite{Hsu}, which leads to a wrong equation of state,
namely that for dust. The second option is the particle horizon
radius. In this case it is impossible to obtain an accelerated
expansion. Only the third choice, the identification of $L$ with
the radius of the future event horizon gives the desired result,
namely a sufficiently negative EoS to obtain an accelerated
universe. However, as soon as an interaction between dark energy
and dark matter is taken into account, the identification of IR
cut-off with Hubble (apparent horizon) radius  can simultaneously
drive accelerated expansion and solve the coincidence problem
\cite{pav1}. In recent years, some new infrared cut-offs have also
been proposed in the literature. In  \cite{Ricci} a HDE model with
Ricci scalar as IR cut-off was proposed, while in \cite{GO} the
authors have added the square of the Hubble parameter and its time
derivative within the definition of HDE. A linear combination of
particle horizon and the future event horizon was also proposed in
\cite{Sad}. In this paper, as system's IR cut-off, we take the
radius of the event horizon measured on the sphere of the horizon
defined as \cite{wang1}
\begin{equation}\label{ldef}
  L=a r(t),
\end{equation}
where $r(t)$ can be obtained from
\begin{equation}\label{rt}
    \int_0^{r(t)}\frac{dr}{\sqrt{1-kr^2}}=\int_0^{\infty}\frac{dt}{a}=\frac{R_h}{a}.
\end{equation}
Solving the above equation for $r(t)$ we obtain
\begin{equation}\label{rtf}
  r(t)=\frac{1}{\sqrt{k}}\sin\left(\frac{\sqrt{k} R_{\rm h}}{a}\right).
\end{equation}
Here $R_{\rm h}$ is the radial size of the event horizon measured
in the $r$ direction. Assuming an interaction between dark energy
and dark  matter, the conservation equations read
\begin{eqnarray}
\dot\rho_m+3H\rho_m&=&Q,\label{conseq1}\\
\dot\rho_D+3H\rho_D(1+w_D)&=&-Q\label{conseq2},
\end{eqnarray}
where $Q=3b^2H(\rho_D+\rho_m)$ is an energy exchange term, $b^2$ is
a coupling constant and $w_D$ is the EoS parameter of the DE
component. Differentiating (\ref{rhopl}) with respect to cosmic time
$t$ and using (\ref{fridomega}), (\ref{ldef}), (\ref{rt}) and
(\ref{rtf}) we have

\begin{equation}\label{rhodot}
    \dot{\rho_D}=H\rho_D\left(\frac{\alpha-2}{\gamma_c}-\alpha\right)\left[1-\frac{1}{c}\sqrt{\frac{\Omega_D}{\gamma_c}}\cos y\right]
\end{equation}
where $y=\sqrt{k}\frac{R_{\rm h}}{a}$. Substituting the above relation in Eq.
(\ref{conseq2}), we obtain \cite{Sheykhi2}
\begin{equation}
w_D=-1+\frac{1}{3}\left(\alpha-\frac{\alpha-2}{\gamma_c}\right)\left[1-\frac{1}{c}\sqrt{\frac{\Omega_D}{\gamma_c}}\cos
y\right]-\frac{b^2(1+\Omega_k)}{\Omega_D}.
\end{equation}
To develop the correspondence between PLECHDE and tachyon field,
we identify $w_D=w_\phi$ and obtain
\begin{equation}
\dot\phi=\sqrt{1+w_D}=\sqrt{\frac{1}{3}\left(\alpha-\frac{\alpha-2}{\gamma_c}\right)\left[1-\frac{1}{c}\sqrt{\frac{\Omega_D}{\gamma_c}}\cos
y\right]-\frac{b^2(1+\Omega_k)}{\Omega_D}}.
\end{equation}
Using $\dot\phi=\phi'H$, we can write
\begin{equation}
\phi'=\frac{1}{H}\sqrt{\frac{1}{3}\left(\alpha-\frac{\alpha-2}{\gamma_c}\right)\left[1-\frac{1}{c}\sqrt{\frac{\Omega_D}{\gamma_c}}\cos
y\right]-\frac{b^2(1+\Omega_k)}{\Omega_D}},
\end{equation}
where the prime denotes derivative with respect to $x=\ln a$. Integrating yields
\begin{equation}\label{phia}
\phi(a)-\phi(a_0)=\int\limits_{a_0}^a\frac{1}{aH}\sqrt{\frac{1}{3}\left(\alpha-\frac{\alpha-2}{\gamma_c}\right)\left[1-\frac{1}{c}\sqrt{\frac{\Omega_D}{\gamma_c}}\cos
y\right]-\frac{b^2(1+\Omega_k)}{\Omega_D}}da,
\end{equation}
where $a_0$  is the  value of the scale factor at the present time
$t_0$. Alternatively we can rewrite Eq. (\ref{phia}) as
\begin{equation}\label{recphi}
\phi(t)-\phi(t_0)=\int\limits_{t_0}^t\sqrt{\frac{1}{3}\left(\alpha-\frac{\alpha-2}{\gamma_c}\right)\left[1-\frac{1}{c}\sqrt{\frac{\Omega_D}{\gamma_c}}\cos
y\right]-\frac{b^2(1+\Omega_k)}{\Omega_D}}dt'.
\end{equation}
Also using relation (\ref{rhophi}), we have
\begin{eqnarray}\label{recv}
V(\phi)=3M_p^2H^2\Omega_D\sqrt{1-\frac{1}{3}\left(\alpha-\frac{\alpha-2}{\gamma_c}\right)\left[1-\frac{1}{c}\sqrt{\frac{\Omega_D}{\gamma_c}}\cos
y\right]-\frac{b^2(1+\Omega_k)}{\Omega_D}}.\nonumber\\
\end{eqnarray}
Therefore, we have established an interacting power-law
entropy-corrected holographic tachyon dark energy model and
reconstructed the potential and the dynamics of the tachyon field.
Such a tachyon scalar field model with  potential (\ref{recv}) and
dynamical equation (\ref{recphi}) can act the role of DE as the
PLECHDE.
\section{Quintessence reconstruction of PLECHDE}  \label{qerec}

We use the term ``quintessence" to denote a canonical scalar field
$\phi$ with a potential $V (\phi)$ that can interact with all
other components only through standard gravity. The quintessence
model is therefore described by the lagrangian
\begin{equation}\label{qelag}
    {\cal
    L}=-\frac{1}{2}g^{\mu\nu}\partial_{\mu}\phi\partial_{\nu}\phi-V(\phi).
\end{equation}
The energy-momentum tensor of quintessence is
\begin{equation}\label{emtensqe}
    T_{\mu\nu}=\partial_{\mu}\phi\partial_{\nu}\phi-g_{\mu\nu}
    \left[\frac{1}{2}g^{\alpha\beta}\partial_{\alpha}\phi\partial_{\beta}\phi+V(\phi)\right].
\end{equation}
In the FRW framework the energy density and pressure of the
quintessence field can be written as
\begin{equation}\label{rhoqe}
    \rho_{\phi}=-T_0^0=\frac{1}{2}\dot{\phi}^2+V(\phi),
\end{equation}
\begin{equation}\label{rhoqe}
    p_{\phi}=T_i^i=\frac{1}{2}\dot{\phi}^2-V(\phi),
\end{equation}
where $\rho_{\phi}$ and $p_{\phi}$ denote the energy density and
pressure, respectively. Using the above relations it can be easily
seen that the kinetic term and the scalar potential are
\begin{equation}\label{ktqe}
    \dot{\phi}^2=(1+w_D)\rho_{\phi},
\end{equation}
\begin{equation}\label{vqe}
    V(\phi)=\frac{1-w_D}{2}\rho_{\phi},
\end{equation}
where $w_D=p_{\phi}/\rho_{\phi}$. Identifying
$\rho_{\phi}=\rho_{D}=\frac{3c^2M_p^2}{L^{2}}\gamma_c$ and using
(\ref{ktqe}) one obtains
\begin{eqnarray}
\dot{\phi}^2&=&3M_p^2H^2\Omega_D\left(\frac{1}{3}\left(\alpha-\frac{\alpha-2}{\gamma_c}\right)\left[1-\frac{1}{c}\sqrt{\frac{\Omega_D}{\gamma_c}}\cos
y\right]-\frac{b^2(1+\Omega_k)}{\Omega_D}\right).\nonumber\\
\end{eqnarray}
Using, $\dot\phi=\phi'H$, we have
\begin{eqnarray}\label{phiprime}
&&\phi(a)-\phi(a_0)=\int_{\ln a_0}^{\ln
a}\\
&&\left(3M_p^2\Omega_D\left(\frac{1}{3}\left(\alpha-\frac{\alpha-2}{\gamma_c}\right)\left[1-\frac{1}{c}\sqrt{\frac{\Omega_D}{\gamma_c}}\cos
y\right]-\frac{b^2(1+\Omega_k)}{\Omega_D}\right)\right)^{1/2}d\ln
a,\nonumber
\end{eqnarray}
where $a_0$ denotes the present value of the scale factor. From Eq. (32) the
scalar potential is obtained as
\begin{eqnarray}
V(\phi)=3M_p^2H^2\Omega_D\left(1-\frac{1}{6}\left(\alpha-\frac{\alpha-2}{\gamma_c}\right)\left[1-\frac{1}{c}\sqrt{\frac{\Omega_D}{\gamma_c}}\cos
y\right]+\frac{b^2(1+\Omega_k)}{2\Omega_D}\right).\nonumber
\end{eqnarray}
One can easily check that the usual holographic quintessence dark
energy model can be retrieved in the limiting case $\gamma_c=1
 (\alpha=\beta=0)$.
\section{K-essence reconstruction of PLECHDE} \label{kessrec}
Quintessence relies on the potential energy of scalar field which
leads to the late time acceleration. It is possible to have a
situation where the accelerated expansion arises out of
modifications to the kinetic energy of the scalar field. K-essence
is characterized by a scalar field with a non-canonical kinetic
energy. The most general scalar-field action which is a function
of $\phi$ and $X=-\dot{\phi}^2/2$ is given by \cite{arm}
\begin{equation}\label{kessaction}
    S=\int d^4x\sqrt{-g}P(\phi,X),
\end{equation}
where the lagrangian density $P(\phi,X)$ corresponds to a pressure
density. According to this lagrangian the energy density and the
pressure can be written as \cite{chiba}
\begin{eqnarray}\label{rhokess}
\rho_\phi&=&f(\phi)(-X+3X^2),\label{rhokess}\\ P_\phi
&=&f(\phi)(-X+X^2).\label{pkess}
\end{eqnarray}
Therefore the EoS parameter of the K-essence will be
\begin{equation}\label{wkess}
    w_K=\frac{P_\phi}{\rho_\phi}=\frac{1-X}{1-3X}.
\end{equation}
To implement the correspondence between K-essence and PLECHDE, we
set $w_K=w_D$ and solve (\ref{wkess}) for $X$. We find
\begin{equation}\label{X}
    X=\frac{1-w_D}{1-3w_D}=\frac{2-\frac{1}{3}\left(\alpha-\frac{\alpha-2}{\gamma_c}\right)\left[1-\frac{1}{c}\sqrt{\frac{\Omega_D}{\gamma_c}}\cos
y\right]+\frac{b^2(1+\Omega_k)}{\Omega_D}}{4-\left(\alpha-\frac{\alpha-2}{\gamma_c}\right)\left[1-\frac{1}{c}\sqrt{\frac{\Omega_D}{\gamma_c}}\cos
y\right]+\frac{3b^2(1+\Omega_k)}{\Omega_D}}.
\end{equation}
Since $\dot{\phi}^2=-2X$ and $\dot{\phi}=H\phi'$ we obtain
\begin{equation}\label{phif}
\phi(a)-\phi(a_0)=\int_{a_0}^a\frac{1}{Ha}\sqrt{\frac{-4+\frac{2}{3}\left(\alpha-\frac{\alpha-2}{\gamma_c}\right)
\left[1-\frac{1}{c}\sqrt{\frac{\Omega_D}{\gamma_c}}\cos
y\right]-\frac{2b^2(1+\Omega_k)}{\Omega_D}}{4-\left(\alpha-\frac{\alpha-2}{\gamma_c}\right)\left[1-\frac{1}{c}\sqrt{\frac{\Omega_D}{\gamma_c}}\cos
y\right]+\frac{3b^2(1+\Omega_k)}{\Omega_D}}}d a.
\end{equation}
\section{Dilaton field reconstruction of PLECHDE} \label{dilrec}
It is quite possible that gravity is not given by the Einstein
action, at least at sufficiently high energy scales. The most
promising alternative seems to be that offered by string theory,
where the gravity becomes scalar-tensor in nature. In the low
energy limit of the string theory, one recovers Einstein gravity
along with a scalar dilaton field which is non-minimally coupled
to gravity \cite{Wit1}. The dilaton field can be used for
explanation the dark energy puzzle and avoids some quantum
instabilities with respect to the phantum field models of DE
\cite{carroll}. The lagrangian density of the dilatonic dark
energy corresponds to the pressure density of the scalar field has
the following form \cite{tsuji}
\begin{equation}\label{dil1}
    P=-X+ce^{\lambda\phi}X^2,
\end{equation}
where $c$ and $\lambda$ are positive constants and
$X=\dot{\phi}^2/2$. Such a pressure (Lagrangian) leads to the
following energy density \cite{tsuji}
\begin{equation}\label{dil2}
    \rho=-X+3ce^{\lambda\phi}X^2.
\end{equation}
The EoS parameter of the dilatonic DE can be written as
\begin{equation}\label{dil3}
    w=\frac{P}{\rho}=\frac{1-ce^{\lambda\phi}X}{1-3ce^{\lambda\phi}X}.
\end{equation}
Seeking for a correspondence between the dilatonic DE and PLECHDE we
set $w=w_D$, where $w_D$ comes from the PLECHDE. Solving this
equation for $cXe^{\lambda\phi}$ we obtain
\begin{equation}\label{dil4}
cXe^{\lambda\phi}=\frac{w_D-1}{3w_D-1}.
\end{equation}
Taking into account that $X=\dot{\phi}^2/2$ we can rewrite Eq.
(\ref{dil4}) as
\begin{equation}\label{dil5}
    \frac{c}{2}\Big(\frac{2}{\lambda}\frac{d}{dt}e^{\lambda\phi/2}\Big)^2=\frac{w_D-1}{3w_D-1}.
\end{equation}
Using $\frac{d}{dt}=H\frac{d}{d \ln a}$ leads
\begin{equation}\label{dil6}
\phi(a)= \frac{2}{\lambda} \ln \Big[e^{\lambda\phi(a_0)/2}+
\frac{\lambda}{\sqrt{2c}}\int_{\ln a_0}^{\ln
a}\frac{1}{H}\Big(\frac{w_D-1}{3w_D-1}\Big)^{1/2}d\ln a\Big],
\end{equation}
which represents the evolution equation of $\phi$. Using the
expression for $w_D$ we can further rewrite the above equation as
\begin{eqnarray}\label{dil7}
&& \phi(a)= \frac{2}{\lambda} \ln \Big[ e^{\lambda\phi(a_0)/2}
\\&&
+\frac{\lambda}{\sqrt{2c}}\int_{\ln a_0}^{\ln a}
\frac{1}{H}\left(\frac{2-\frac{1}{3}\left(\alpha-\frac{\alpha-2}{\gamma_c}\right)\left[1-\frac{1}{c}\sqrt{\frac{\Omega_D}{\gamma_c}}\cos
y\right]+\frac{b^2(1+\Omega_k)}{\Omega_D}}{4-\left(\alpha-\frac{\alpha-2}{\gamma_c}\right)\left[1-\frac{1}{c}\sqrt{\frac{\Omega_D}{\gamma_c}}\cos
y\right]+\frac{3b^2(1+\Omega_k)}{\Omega_D}}\right)^{1/2}d\ln
a\Big]. \nonumber
\end{eqnarray}
\section{concluding remarks}\label{con}
A possibility to explain the origin of the black hole entropy is
the entanglement of quantum fields between in and out the horizon
\cite{Sau}. It was shown \cite{Sau} that the black hole entropy is
proportional to the horizon area when the field is in its ground
state, while a correction term proportional to a fractional power
of area results when the field is in a superposition of ground and
excited states. For large horizon areas, these corrections are
relatively small and the area law is recovered. Taking into
account the correction to area law the HDE density is modified as
well. Based on this, a so-called PLECHDE was  proposed recently
\cite{Sheykhi2} to explain the acceleration of the cosmic
expansion. On the other hand, there are many different models of
DE in the context of a scalar field such as tachyon, quintessence,
K-essence and dilaton fields. One exciting question is that
whether a scalar field model can act as a dark energy model, while
they have apparently distinct instinct. In this paper a connection
between the  scalar field models and the PLECHDE has established.
As a result and using the proposal (14) we have reconstructed the
potentials as well as the evolutionary forms of scalar fields. In
section II we presented a tachyon field version of the PLECHDE.
Then, we reconstructed the potential term and the dynamical
equation governing the evolution of the tachyon field. In section
\ref{qerec} we established a correspondence between the PLECHDE
and a quintessence model. The potential for the K-essence
holographic correspondence was derived in section IV. We also
considered the dilaton condensate without potential, and used the
correspondence to find the evolutionary form of the scalar field.
In the limiting case $\gamma_c=1 (\alpha=0=\beta)$, where the
power-law correction becomes trivial, all our results reduce to
their corresponding expressions in standard HDE scalar field
models.

\acknowledgments{ This work has been supported financially by
Research Institute for Astronomy and Astrophysics of Maragha, Iran.}

\end{document}